\documentclass[aps,twocolumn,preprintnumbers,amsmath,amssymb]{revtex4-1}
\usepackage{graphicx}
\usepackage{subfigure}
\usepackage{epsfig}
\usepackage{dcolumn}
\usepackage{bm}
\usepackage{ulem}
\usepackage{color}
\usepackage{float}
\usepackage[colorlinks]{hyperref}
\hypersetup{
	citecolor = cyan
}

\def\be{\begin{equation}}       \def\ee{\end{equation}}
\def\bea{\begin{eqnarray}}      \def\eea{\end{eqnarray}}
\def\ba{\begin{array}}
\def\ea{\end{array}}
\def\bnum{\begin{enumerate} }
\def\enum{\end{enumerate}}

\def\=>{\Rightarrow}
\def\>{\rightarrow}

\def\eye2{Fathbb{I}}

\renewcommand{\>}{\rangle}

\newcommand{\eq}[2]{
	\begin{equation}
	#1 \label{#2}
	\end{equation}
}

\newtheorem{problem}{Problem}

\begin{document}
\title{Classification on the Computational Complexity of Spin Models}
\author{Shi-Xin Zhang}

\email{zsx16@mails.tsinghua.edu.cn}
\affiliation{Institute for Advanced Study, Tsinghua University, Beijing 100084, China}

\begin{abstract}
In this note, we provide a unifying framework to investigate the computational complexity of classical spin models and give the full classification on spin models in terms of system dimensions, randomness, external magnetic fields and types of spin coupling. We further discuss about the implications of NP-complete Hamiltonian models in physics and the fundamental limitations of all numerical methods imposed by such models. We conclude by a brief discussion on the picture when quantum computation and quantum complexity theory are included.
\end{abstract}

\date{\today}
\maketitle

\textit{Introduction:}
Computational complexity classes are very important tools in computer science to characterize the hardness of problems \cite{Sipser2006}. After the original work of Cook\cite{Cook71}, NP-completeness (NPC)\cite{NPbook} and in general the notion of complete problems of corresponding complexity classes have become the dominant approach for addressing how hard a problem is.
For decision problems in nondeterministic polynomial time (NP), namely decision problems that can be solved in polynomial time on a nondeterministic Turing machine, they can be classified as  polynomial time (P), NP-complete or NP-intermediate (neither in NPC nor in P). In this work, we take the conjecture $\mathrm{P}\neq \mathrm{NP}$ throughout\cite{Aaronson}, implying NP-complete problems are intractable. Surprisingly, most of the common NP problems are either NP-complete or in P and there are fewer NP-intermediate problem candidates than na\"{i}ve thought \footnote{factorization is one of the most famous NP-intermediate candidates.}. This fact shows the deep structure of NP problems and the universality of  NP-completeness language. 

Barahona\cite{Barahona1982} introduced the NP-completeness notion into statistical mechanics by considering the ground state decision problems and gave the proof of NP-completeness on two types of classical spin models. Furthermore, following previous works on exactly solving Ising models and dimer models\cite{Onsager1944, Kasteleyn1961, Kasteleyn1963, Fisher1966, Bieche1980}, Barahona gave a general polynomial-time algorithm which can exactly solve the ground state energy for all 2D spin models without external magnetic field on arbitrary lattice. Therefore, it is clear the classical spin models are totally different in terms of computational complexity. In this work, we will give a full classification on the \textit{hardness} of spin models: some of them are claimed in P by directly providing polynomial-time algorithms and the others are shown as NP-complete by the reduction proofs from 3SAT. The framework for NPC proof in this work is inspired by the results on the computational complexity of random field Ising model\cite{AnglesdAuriac1985}.

 Notion of NP-completeness in physics, as an intrinsic property of some statistical models,  is not only of academic interest, but also plays an vital role when such physics systems are numerically investigated in practice. For example, there are various works dealing with the \textit{hardness} of specific numerical methods with the help of models in NP(QMA)-complete complexity classes\cite{Troyer2005, Schuch2009}. In this work, we elaborate on this idea and show that the fundamental limitations of numerical approximation are universal to all numerical simulation schemes due to the existence of NP-hard models. This fact brings invaluable insight into the numerical study on physics systems in general. Efficient numerical schemes often fail in some models (namely the time to get reasonable approximation results scales exponentially with the system size), and in such cases we often attribute the failure to the algorithm itself and try to update the algorithm or replace it with other schemes. However, when various numerical schemes fail on the same model due to seemingly different reasons, it is highly possible that the hardness is not due to the drawbacks of individual numerical methods,  but from the model itself,  indicating that there is no efficient numerical simulation method at all. We also discuss how the above picture remains the same when quantum computation is allowed.

\textit{Notations:} 
A decision problem is a problem that can be posed as a yes-no question of the input values. The set of decision problems can be solved in polynomial time on a deterministic Turing machine in terms of the input size is P.  As for NP class, there are two equivalent definitions. The sets of decision problems can be solved in polynomial time on a nondeterministic Turing machine is NP, or the sets of problems that the yes answer instances can be verified given a proof in polynomial times is NP. It is obvious $P\subseteq NP$, as P problems require null proofs. Problem reduction is defined as a function $f$, such that for any legal input $x$ of problem A, we have $f(x)$ as input of problem B and $A(x)=1\iff B(f(x))=1$. If such a problem reduction $f$ can be carried out in polynomial time by a Turing machine, we have $A<_P B$. Intuitively, problem A is no harder than B, since solution to B also solves problem A. If $A<_P B$ and $B<_P A$, problem A and B are said to be computationally equivalent, written as $A=_P B$. If $\forall Q\in NP, Q<_P R$, we call such problems R NP-hard. A problem is NP-complete if it is both in NP and NP-hard. It is conjectured that $\mathrm{P}\neq \mathrm{NP}$. 
 For more formal definitions and discussions on the computational complexity classes, please refer to \cite{Sipser2006}.

We use standard conventions on the definition of a graph. A graph $G=(V,E)$ is composed of vertices $V$ and edges $E$ connecting two vertices. If the edge has (no) direction, the graph is called (un)directed graph. $E(G)$($V(G)$) corresponds for the set of edges(vertices) of the graph $G$.  The size of the two sets are   $|E(G)|=m$, $|V(G)|=n$. We can further attach one real value for each edge as $w(E)$, such a graph is called weighted graph. The degree of a vertex is the number of edges incident to the vertex, and a graph with all vertices of degree $3$ is called cubic graph. A planar graph is a graph that can be embedded on the plane without any edges crossing, and the else are nonplanar graph. A plane graph is a planar graph that has already been embedded on the plane, there may be more than one plane graph (choice of planar embedding) for a planar graph. The dual graph of a plane graph $G$ is a graph that has a vertex for each face of $G$. The dual graph has an edge whenever two faces of $ G$ are separated from each other by an edge. Thus, each edge $e$ of $G$ has a corresponding dual edge, whose endpoints are the dual vertices corresponding to the faces on either side of $e$. 

A cut is a partition of the vertices of a graph into two disjoint subsets (one of the subsets is defined as  a base set $S$). Any cut $S$ determines a cut-set $\mathrm{CUT}(S)$, the set of edges that have one endpoint in each subset of the partition.  The size of the cut-set is defined as number of edges in the cut-set for unweighted graph, which is also called simple cut. For weighted graph, the total weight of the cut-set is the sum of all weights of edges from the cut-set. The simple MAX-CUT problem is defined as:
\begin{problem}
Given the unweighted graph G and an integer k, is there any vertices set $S$, such that $|\mathrm{CUT}(S)|\geq k$?
\end{problem}
The simple MAX-CUT problem is NP-complete even on general cubic graph \footnote{[NP16] comment in \cite{NPbook},  transformation from NAE-3SAT}, while simple MAX-CUT problem is in P on planar graph\cite{Hadlock1975}. Similarly we can define MAX-CUT problems on weighted graphs using the notation of $w(\mathrm{CUT})=\sum_{e\in \mathrm{CUT}}w(e)$ and one can also define MIN-CUT problems similarly. 

The MAX-CUT problem separating two given vertices, is defined as: 
\begin{problem}
Given graph $G$, two vertices $s,t$ on the graph and the fixed value $k$, is there any base set $S$, such that $s\in S$, $t\notin S$ and $w(\mathrm{CUT(S)})\geq k$?
\end{problem}
 It can be shown that MAX-CUT problem and MAX-CUT problem separated by two given vertices are computationally equivalent. MAX-CUT $<_P $ MAX-CUT separating by two given vertices: Add two virtual vertices $s,t$ on the original graph and connect $s,t$ to all other vertices with weights $0$, and thus the solution for  MAX-CUT separating $s$ and $t$ is the solution to MAX-CUT on original graph. MAX-CUT separating by two given vertices $<_P$ MAX-CUT: Add an edge connecting the two give vertices and assign weight $w=\sum_{e\in E(G)} |w_e|$ on this new edge. By this operation the two given vertices are forced in different sides of the cut and thus a solution for MAX-CUT separates the two given vertices on the different sides of the cut. 

A spin model is defined by a Hamiltonian $H$ on a lattice graph $G$. The Hamiltonian is of the form 
\begin{equation}
	H=\sum_{\langle ij\rangle\in E(G)}-J_{ij}S_iS_j+\sum_{i\in V(G)} h_i S_i,
\end{equation}
where the spin freedoms $S_i$ live on the vertices of the lattice graph which can take value $S_i=\pm 1$.  Spin couplings $J_{ij}$  are defined on edges of lattice graph $G$, while external magnetic fields $h_i$  are defined on vertices. The ground state energy of the model is defined as the minimal value of total  energy among all spin configurations, namely
\eq{E_0(H)=\min_{\{S_i\}}H(\{S_i\}),}{arg2}
 where $\{S_i\}$ take values from configuration space with size $2^n$. 
 
 The groud state energy decision problem is defined as:
 \begin{problem} Given the input of the spin model (including the lattice graph $G$ and all parameters in the Hamiltonian $J_{ij}$, $h_i$) and a fixed value $E_k$, is the ground state energy of the model $E_0(H)\leq E_k$? 
 \end{problem}
 This question is formulated as a decision problem and it is obviously in NP since a nondeterministic Turing machine can naturally explores all spin configurations  at the same time by different branches and make the comparison at the end of each branch. Therefore, the remaining task is to classify all types of spin models into P and NP-complete classes based on their ground state energy decision problems.
 
 We divide spin models in terms of the following conditions. (1) Dimension: the dimension of the model is related with the underlying lattice graph. For planar graph, the dimension is $2$, while for nonplanar graph, the dimension is $3$ \footnote{It is obvious to observe that all graphs can be embedded in 3D without edges crossing}. In the physics language, $D=3$ in terms of lattice graph corresponds to $D\geq 3$ in real systems. (2) Spin-spin coupling sign: If $\forall e\in E(G), J_{e}\geq0$, we call such model ferromagnetic (FM) models. Otherwise for models $\exists e\in E(G), s.t.~ J_{e}<0$, they are said to be antiferromagnetic (AFM) models.  (3) Randomness: If all spin-spin couplings and external fields are constants as $\forall e\in E(G), v\in V(G), J_{e}=J, h_{v}=H$, such models are uniform. Instead, if parameters in the Hamiltonian are allowed to taken in several values or a continuum of parameter windows, the corresponding models are called random models. The concept of couping signs can combine with randomness. For example, Hamiltonians of random FM models satisfy $\forall e \in E(G), J_{e}\geq 0$.  (4) External magnetic fields: If $\forall v \in V(G), h_{v}=0$, such systems are said to be without external fields and otherwise the models have external fields.

\textit{Classifications:}
Based on the four conditions of spin models, we have $2^4=16$ types of spin models to be classified. We first introduce a unified mapping (FM transformation) between ground state energy problems in spin models and cut-related problems on graphs. The mapping from the spin models to graph problems are defined as below. Given a Hamiltonian and the underlying lattice graph $G$, we first decorate edge $e$ with corresponding $J_e$ value as weights . If the model has external fields $h_{v}\neq 0$, we add two vertices $s,t$  and connect edges from $s$ ($t$) to all vertices with positive (negative) external field $h_{v}>0$ ($h_{v}<0$) on the lattice graph. Moreover, we assign the newly added edges $e=(s,v)$ ($e=(t,v)$) of weights $w(e)=|h_v|=h_v$ ($w(e)=|h_v|=-h_v$). The decorated weighted graph after FM transformation is denoted by $G'$. The inverse transformation from any weighted graph $G'$ to a spin model $H$ with underlying lattice graph $G$ is obvious. Just define the Hamiltonian on the graph $G'$ without external field and with $J_{e}=w(e)$. Note the mapping can be easily carried out in polynomial time.

Given a configuration of spins $\{S_i\}$, consider the set of vertices $S$ for cut on $G'$ including all vertices with spin up and vertex $s$, i.e. $S=\{v|S_v=1 \}\bigcup s$.  It is obvious to see that the total energy of the system is given by
\eq{H(\{S_i\})=-E_0+2w(\mathrm{CUT}(S(\{S_i\}))),}{htocut}
where $E_0$ is a constant independent of the spin configurations which can be obtained in polynomial time. To solve the ground state energy problem, the only task is to minimize the cut part in \eqref{htocut}, which is the problem of min cut separating two given vertices.

Similarly, we can also define the AFM transformation. In such a mapping, the weight of edges on $G'$ is $-J_{e}$ and the base set $S$ for the cut is the union of spin up vertices and $t$ vertex. Under this type of transformation, the energy of the system is given by $E_0'-2w(\mathrm{CUT}(S))$. Therefore, by AFM transformation, the problem of solving ground state energy is equivalent to the problem of max cut separating two given vertices which is further equivalent to the max cut problems.

In the case of FM transformation, this ground state energy problem is reduced to MIN-CUT separating by two vertices as we already shown. If all weights of edges are positive, according to min-cut max-flow theorem, the problem can be transformed into max flow problem\cite{Cormen2009, Goldberg2014} on a graph which is known in P\footnote{If one only cares about complexity classes, then max flow can be shown in P by reformulate the problem as  linear programming which is guaranteed in P\cite{Khachiyan1979}}. This is the case when the models have positive spin couplings, i.e. FM models. The reduction chain for these cases are ground state energy problem for FM spin models $<_P$ MIN-CUT separating two given vertices $<_P$ max-flow problem (P). Namely for all models with FM couplings in 2D and 3D, the ground state energy can be solved in polynomial time with or without external fields or randomness. Specifically, the ground state energy can be derived in $O(nm)$ time by the state-of-the-art max-flow algorithm\cite{Orlin2013}. For short-ranged couplings (sparse lattice graph $m=O(n)$), it is shown that the ground state energy problem can be exactly solved in $O(n^2/\log n)$ time where $n$ is the system size. As for the approximation solution, there exists an almost linear time algorithm for max flow with precision $\varepsilon$ in $O(m^{1+O(1)}\varepsilon^{-2})$ time\cite{Kelner2013}.

For AFM coupling spin models, recall the result from Barahona\cite{Barahona1982}, which provided the polynomial algorithm for AFM spin model in 2D without external fields. By utilizing the one-to-one correspondence between spin configurations and sets of unsatisfied edges, we can transform the ground state energy problem on graph $G$ into the the max-weight perfect matching (MWPM) problem on graph $G^\star$ which is a modified graph from the dual graph of $G$.  The algorithm exploits the famous blossom algorithm finding max-weight perfect matching on the graph\cite{Edmonds1965, Cunningham1978, Galil1986}. The complexity of the state-of-the-art algorithm for this task is known as $O(mn+n^2\log n)$\cite{Gabow2016}. Besides, if all the spin-spin couplings are integers with the integer absolute value bounded by $W$,  the MWPM algorithm can be further improved to $O(m\sqrt{n}\log nW)$\cite{Duan2018}. For example, a model where short-ranged spin couplings taking values from $\pm 1, 0$ can be solved exactly in $O(n\sqrt{n}\log n)$ time where $n$ is the system size. From our unified picture of max cut analysis, the transformation algorithm by Barahona can be understood as a strong version of the proof that planar MAX-CUT is in P, i.e. MAX-CUT problem with positive and negative weights on planar graph is in P. Hence the reduction chain can be understood as 2D AFM spin model without external fields $<_P$ MAX-CUT on planar graph with weights of both signs (P).

All classical spin models whose ground state energy problems are in P can be reduced to either max-flow problem or max weight perfect matching problem. The highlight is that these two problems are probably the most famous and representative combinatoric problems in the field of graph theory. Meanwhile they have been extensively studied for several decades with various cutting-edge results both for efficient exact and approximation approaches. Therefore, the state-of-the-art algorithms for these two problems might provide powerful new tools for the study on spin glass systems.

There are already $10$ classes of spin models out of $16$ are claimed in P. For the remaining $6$ classes of model, we claim they are all NP-completeness in terms of ground state energy problem. Since trivially 3D spin models with uniform AFM couplings without external fields $<_P$ counterparts with random AFM couplings; 2D spin models with uniform external field and spin couplings $<_P$ counterparts in 3D or random version models in 2D (all the above reduction are in terms of ground state energy decision problems), we only need to prove NP-completeness for two classes of models. Barahona \cite{Barahona1982} has shown that 2D uniform AF coupling spin models with uniform external field is NP-complete by the reduction from max independent set problem on planar cubic graphs \footnote{[GT20] comment in \cite{NPbook}, transformation from planar 3SAT}. Based on this known results, we can provide a unified understanding in hindsight by the AFM transformation. Since this problem is computationally equivalent to the MAX-CUT problem on graph $G$, where $G$ is a positive weighted cubic planar graph plus two extra vertices, and the sets of vertices connected from the two extra vertices have no overlap. This fact implies that even a small deviation from planar graph can make MAX-CUT problem from P to NP-completeness.

For the 3D uniform AFM spin models without external field, we apply the AFM transformation on such model, and it gives the reduction from MAX-CUT problem on general graph to the ground state energy problem in this class of models. Therefore, it is straightforward to conclude that the ground state energy problem of 3D uniform AFM spin models is NP-complete on cubic graph.

We summarize the full classification on the computation complexity of spin models by Table \ref{resulttable}. We omit random AFM with external fields line, since they are trivially in NPC as their uniform counterparts are already in NPC. And all NPC problems in the table are still in NPC even restricted to cubic graphs inputs. The unified framework to understand the classification is provided by the FM/AFM transformation we discussed, and the investigation on the complexity classes of corresponding MAX(MIN)-CUT problems on some subsets of (un)weighted graph. The general guiding principles include: on positive weighted graphs, (1) MAX-CUT is in NPC; (2) MIN-CUT is in P (max flow); (3) MAX-CUT is in P if restricted to planar graph (MWPM).

\begin{table}[]
	\centering
	\caption{Summary of the classification on the computational complexity of spin models.}
	\label{resulttable}
	\begin{tabular}{|c|c|c|}
		\hline
Spin	Models	& 2D                        & $\geq$3D                     \\
		\hline
		FM                    & P             & P           \\
		\hline
		uniform AFM, no fields & P                & NPC   \\
		\hline
		random AFM, no fields  & P                 & NPC  \\
		\hline
uniform AFM, external fields    & NPC & NPC                    \\ 
	\hline                
	\end{tabular}
\end{table}

There are several observations immediately from the classification results. Firstly, the AFM models are no easier than their FM counterparts. This fact is intuitive and can be justified by the more complex energy landscape for AFM models. Secondly, randomness brings no further hardness at least in spin models investigated here. Namely, if a uniform spin model is in P then its counterpart with random couplings is also in P. And if a random spin model is NP-complete, the uniform counterpart is also in NPC. This fact may provide new insights into our understanding on randomness and spin glasses. Besides,  it is worth noting that the computational hardness has nothing to do with whether the model has a finite temperature phase transition.

\textit{Discussions:}
The NP-completeness of ground state energy problems imposes strong limitations on the numerical methods even for efficient approximations.  A reasonable efficient approximation is the algorithm giving solutions $E_1$ as an approximation of the true ground state energy $E_0$ with precision $\varepsilon$ ($|E_1-E_0|\leq \varepsilon E_0$) in polynomial time with the input size $n$ and precision $1/\varepsilon$ as $O(n^\alpha\varepsilon^{-\beta})$. This efficient approximations is one of the basic requirements for any reasonable numerical approaches and the concept is summarized as fully polynomial time approximation scheme (FPTAS).

Since all the NP-complete spin models are intrinsically hard even when all couplings are restricted to $\pm 1$, they are denoted by NP-complete in the strong sense \cite{Garey1978}. It is easy to prove that there is no FPTAS for strong NP-complete problems with optimal aim of integer values. Take the uniform AFM 3D spin models for an example, suppose all the couplings are $J_e=1$, the total energy of the system is integer valued. If there is FPTAS for ground state energy problem of such models, we can do the approximation algorithm with precision $\epsilon = 1/(2E_0)$, such an approximation leads to the approximation solution $|E_1-E_0|\leq 1/2$. Since the energy spectrum is integer valued, $E_0=E_1$, and thus we obtain the exact results by approximation scheme with the time $O(n^{\alpha}m^{\beta})$. Therefore we cannot calculate the approximate ground state energy efficiently  unless $\mathrm{P}=\mathrm{NP}$.

From the above analysis, NP-complete physics systems give fundamental limitations to all numerical methods in general since no efficient approximations are allowed for such systems. An example is Markov chain Monte Carlo (MCMC)\cite{Binder1995}. It is known that there exists efficient update algorithms for 3D FM spin models while the slowing down and the exponential increase of autocorrelation time is unavoidable in general 3D AFM spin systems. This fact in classical MCMC is well explained in the framework of computational complexity analysis and shows how the complexity theory plays a vital role in understanding the behavior of numerical methods. One may try to improve the cluster update algorithm expecting that the slowing down problem can be avoided in general. Since the intrinsic hardness is in the model itself, this task is as hard as proving $\mathrm{P}=\mathrm{NP}$.

The classical models also add limitation to numerical methods for quantum systems. For example, we can use quantum Monte Carlo (QMC) to solve 3D AFM spin models by treating the spin as Pauli matrix instead of a number. Since this family of models is NP-complete, QMC must fail in polynomial times and the obstacle is nothing but notorious sign problem\cite{Loh1990, Li2018}. The fact shows that sign problem of QMC cannot be systematically solved as it is NP-hard\cite{Troyer2005}. We can go a step further, by calculating the same model both in classical MCMC and QMC as mentioned above, the algorithms fail due to different reasons: slowing down or sign problem. In some sense, the two obstacles of MCMC are two sides of the same coin. It  is of the same difficulty to systematically avoid them. Based on the discussion on fundamental limitations of numerical approaches, we demonstrate the new understanding on why numerical methods become inefficient sometimes. It may be useless to propose any fancy improvements on numerical approaches to solve certain models because the models are \textit{hard} by their own and the inefficient bottlenecks in each numerical schemes are just reflections of NP-completeness from the models.

The NP-complete problems always come with inputs encoding scalable infinite freedoms. In this work, the freedoms are encoded in the input lattice graphs. This is because all problems in NP can be reduced to NPC problems, and the encoded large freedoms correspond to each NP problems. In other words, it is hard to believe one can solve all problems in NP by a problem with one integer input. So we can not talk about the NP-completeness of very specific models. For example,  the input for AFM model with uniform fields and uniform couplings on some fixed lattice has only three integers: size $n$, coupling $J$ and external field $H$. Ground state energy problem of such specific model is hard to describe in the context of NP-completeness, since it is of little hope that every problem in NP can be easily reduced to a problem with three natural freedoms. There is no good way to characterize the complexity of such problems which is too specific to be NP-complete but might also hard at the same time.

On the other hand, the subset of NP-complete problems are not necessarily hard. They can be in P as well. For example, 3D uniform AFM model without external field on arbitrary graph is NP-complete. But for its subset where the given graph is 3D grids, the ground state configuration and energy value is obvious in P.  This fact leaves another possibility how we can study NP-complete systems. Although we have no efficient numerical approach on NPC models, we can still compute the subset we are interested by showing the subset of the models we care about is actually in P. Namely, one can always try to make a finer classification on the computation complexity of models. Not only a subset of problems from NP might be in P, a collection of NPC problems may also become easy. For example, although the ground state energy for each disorder configuration is NP-complete problem, the disorder averaged results are not necessarily NP-complete due to the possible emergent structure of disorder average.

Finally, we give a brief discussion on the picture when quantum computation and quantum complexity theory are included. Firstly, it is widely believed that a quantum computer also can't solve NP-complete problems efficiently (BQP cannot cover NPC). Therefore, the whole discussion in this work may still hold even if quantum computer are used to carry out numerical simulations on NP-complete physical system. As for the complexity classification on quantum models, it is more suitable to use the QMA-complete\cite{Kitaev2002} class to define hard problems. QMA is the quantum counterpart of NP class, and thus QMA-complete is no easier than NP-complete problems. Various quantum Hamiltonians\cite{Bookatz2014} have been recognized as QMA-complete in terms of ground state energy promise problems since the initial work for 5-local Hamiltonians\cite{Kitaev2002}. All the general ideas and insights in the above discussions has no substantive change considering QMA-complete problems. For example, the fundamental limitations of density functional theory(DFT) is shown by QMA-complete quantum models\cite{Schuch2009}. Perhaps the most amazing conjecture from NP(QMA)-complete models is that the ground state energy is inaccessible in polynomial time even in experiments. It is the spirit of the strong version of Church-Turing thesis which states that no nature process can be faster than a Turing machine in terms of computation complexity\cite{Arute2019}\footnote{The strong version of Church-Turing thesis is claimed to be violated by \cite{Arute2019}. }.

In conclusion, we give a full classification on the computational complexity of classical spin models by a unifying framework of FM/AFM transformations. The ground state energy problems for spin models are related with some kinds of MAX(MIN)-CUT problems under the transformations. Furthermore, we discuss in detail on how to correctly understand NP-complete problems in physics and how NP-completeness impose fundamental limitations on numerical approximations. The effectiveness of our understanding remains when quantum complexity is considered.

\textit{Acknowledgment:} We thank Hong Yao and Zi-Xiang Li for useful discussions.

\end{document}